\documentclass[%
preprint,
showpacs,preprintnumbers,
 amsmath,amssymb,
 aps,
]{revtex4-1}

\usepackage{graphicx}
\usepackage{dcolumn}
\usepackage{bm}

\graphicspath{{figure/}}
\begin{document}


\title{Backbendings of Superdeformed bands in $^{36,40}$Ar}

\author{Xu-Hui Xiang}

\author{Xiao-Tao He}
\email{hext@nuaa.edu.cn}
\affiliation{College of Material Science and Technology, Nanjing University of Aeronautics and Astronautics, Nanjing 210016, China}

\begin{abstract}
Experimentally observed superdeformed (SD) rotational bands in $^{36}$Ar and $^{40}$Ar are studied by the cranked shell model (CSM) with the paring correlations treated by a particle-number-conserving (PNC) method. This is the first time the PNC-CSM calculations are performed on the light nuclear mass region around $A=40$. The experimental kinematic moments of inertia $J^{(1)}$ versus rotational frequency are reproduced well. The backbending of the SD band at frequency around $\hbar\omega=1.5$ MeV in $^{36}$Ar is attributed to the sharp rise of the simultaneous alignments of the neutron and proton $1d_{5/2}[202]5/2$ pairs and $1f_{7/2}[321]3/2$ pairs, which is the consequence of the band crossing between the $1d_{5/2}[202]5/2$ and $1f_{7/2}[321]3/2$ configuration states. The gentle upbending at the low frequency of the SD band in $^{40}$Ar is mainly effected by the alignments of the neutron $1f_{7/2}[321]3/2$ pairs and proton $1d_{5/2}[202]5/2$ pairs. The PNC-CSM calculations show that besides the diagonal parts, the off-diagonal parts of the alignments play an important role in the rotational behavior of the SD bands.
\end{abstract}


\maketitle


\section{\label{sec:level1}INTRODUCTION}

Since the first observation of the superdeformed rotational band in $^{152}$Dy~\cite{TwinP1986_PRL57}, numerous superdeformed bands have been discovered in the "traditional" superdeformed regions of mass numbers $80, 130, 150$ and $190$. The latest superdeformed archipelago has been found in the light mass region around $A=40$. High spin states of the 
superdeformed rotational bands have been successfully populated in experiment for $^{36}$Ar~\cite{SvenssonC2000_PRL85,SvenssonC2001_PRC63}, $^{40}$Ar~\cite{IdeguchiE2010_PLB686}, $^{40}$Ca\cite{IdeguchiE2001_PRL87} and $^{44}$Ti\cite{OLearyC2000_PRC61}. Most interestingly, these nuclei are magic or near-magic systems, whose ground states are corresponding to a spherical shape. This exotic shape coexistence phenomenon provides an ideal test ground of theoretical models. 

Many microscopic descriptions of these bands have been performed, like Cranked Nilsson-Strutinsky (CNS)~\cite{SvenssonC2000_PRL85}, Shell Model (SM)~\cite{SvenssonC2000_PRL85,PovesA2004_NPA731,CaurierE2005_PRL95,CaurierE2007_PRC75}, Cranked Relativistic Mean-Field (CRMF)~\cite{IdeguchiE2001_PRL87}, Hartree-Fock BCS with Skyrme interaction SLy6~\cite{BenderM2003_PRC68}, Angular Momentum Projected Generator Coordinate (AMP-GCM) method with the Gogny force D1S~\cite{Rodrguezguzmn2004_IJoMPE13}, Projected Shell Model (PSM)~\cite{LongG2001_PRC63,YangY2015_eprint}, Multidimensionally Constrained Relativistic Mean Field (MDC-RMF)~\cite{LuB2014_PRC89}, Antisymmetrized Molecular Dynamics (AMD)~\cite{Kanada-EnyoY2005_PRC72,KimuraM2006_NPA767,TaniguchiY2007_PRC76,TaniguchiY2010_PRC82}, Cluster models~\cite{SakudaT2004_NPA744_Ar36} and Cranked Hartree-Fock-Bogoliubov (CHFB) etc. Each of these models can give a good description of the certain aspects of these superdeformed nuclei under certain assumptions. Therefore, comprehensive understanding of the superdeformed nuclear structure of these magic or near magic nuclei needs a complementary investigations of different models. Among these models, as it states in Refs.~\cite{CaurierE2005_PRL95,CaurierE2007_PRC75}, the interacting shell model, when affordable, is a prime choice. However, to carry out a practical shell model calculations of $^{36}$Ar, the $1d_{5/2}$ orbital had to be excluded from the $sd$-$pf$ shells space~\cite{SvenssonC2000_PRL85}. Recently, shell model calculations were performed on $^{46}$Ti where a limited configuration space consisting of $1d_{3/2}$ and $1f_{7/2}$ orbitals is constructed. But the full $sd$-$pf$ calculations are still not possible~\cite{MedinaN2011_PRC84}. Therefore, as the full $sd$-$pf$ shell model calculations are difficult in the $A=40$ mass region so far, to test an efficient shell model truncation scheme for the well-deformed nuclei in light mass region is necessary.

Cranked shell model has been proved to be a powerful tool to study the nuclear collective rotation of the most of areas in the nuclear chart. However, up to now, there is no cranked shell model calculation being performed on the SD bands around $A=40$ region. For the first time, we perform the cranked shell model calculations with the pairing treated by the particle-number conserving (PNC-CSM) method on the SD bands in such a light nuclear mass region. The PNC-CSM method is proposed to treat properly the pairing correlations and blocking effects. It has been applied successfully for describing the properties of normal deformed nuclei in A$\sim$170 mass region~\cite{ZengJ1994_PRC50,WuC1991_PRC44,ZengJ2002_PRC65,LiuS2002_PRC66,LiuS2004_NPA735,ZengJ2001_PRC63}, superdeformed nuclei in A$\sim$150, 190 mass region~\cite{WuC1992_PRC45,LiuS2002_PRC66a,LiuS2004_NPA736,ZengJ1991_PRC44,HeX2005_NPA760}, high-K isomeric states in the rare-earth and actinide mass region~\cite{LiB2013_CPC37,ZhangZ2009_PRC80,ZhangZ2009_NPA816,FuX2013_PRC87} and recently in the heaviest actinides and light superheavy nuclei around Z$\sim$100 region~\cite{LiY2016_SCPMA59,ZhangZ2013_PRC87,HeX2009_NPA817}. In contrast to the Bardeen-Cooper-Schrieffer (BCS) or Hartree-Fock-Bogolyubov (HFB) approach, the Hamiltonian is diagonalized directly in a truncated Fock space in the PNC method~\cite{ZengJ1994_PRC50a,ZengJ1983_NPA405}. Therefore, particle number is conserved and Pauli blocking effects are taken into account exactly. 

In the present work we focus on the case of $^{36}$Ar and its heavier isotope $^{40}$Ar, of which superdeformed rotational bands have been established up to the high spin. The present PNC-CSM calculations can reproduce the experimental extracted moment of inertia within an acceptable deviation. This indicates that the PNC-CSM method is an appropriate approach in the light mass region around $A=40$. The observed backbendings of the rotational bands can be understand in the PNC-CSM frame work as the band crossing between the $[321]3/2$ and $[202]5/2$ configuration bands (for both of the neutron and proton). Note that the Nilsson $[321]3/2$ and $[202]5/2$ levels stem from the spherical $1f_{7/2}$ and $1d_{5/2}$ orbitals, respectively. Therefore the effect of the $1d_{5/2}$ orbital is nontrivial on these rotational SD bands.   

\section{\label{sec:level2}THEORETICAL FRAMEWORK}

The cranked shell model Hamiltonian of an axially symmetric nucleus in the rotating frame reads,
\begin{equation}
 H_{\text{CSM}}=\sum_{n}(h_{\text{Nil}}-\omega j_{x})_{n}+H_{\text{P}},
\end{equation} 
where $h_{0}(\omega)=h_{\textrm{Nil}}-\omega j_{x}$ is the single-particle part with $h_{\textrm{Nil}}$ being the Nilsson Hamiltonian~\cite{NilssonS_DMFM29,NilssonS1969_NPA131} and $-\omega j_{x}$ being the Coriolis force with the cranking frequency $\omega$ about the $x$ axis. 
The cranked Nilsson orbitals are obtained by diagonalizing the single-particle Hamiltonian $h_{0}(\omega)=h_{\textrm{Nil}}-\omega j_{x}$. $H_{\text{P}}=H_{\text{P}}(0)+H_{\text{P}}(2)$ is the pairing including monopole and quadrupole pairing correlations. The corresponding effective pairing strengths $G_0$ and $G_2$ are connected with the dimension of the truncated Cranked Many-Particle Configuration (CMPC) space\cite{WuC1989_PRC39} in which $H_{\text{CSM}}$ is diagonalized. In the following calculations, the CMPC space for $^{36,40}$Ar is constructed in the $N = 0\sim4$ major shells for both of neutrons and protons. By taking the cranked many-particle configuration truncation (Fock space truncation), the dimensions of the CMPC space are about 500, the corresponding effective monopole and quadrupole pairing strengths are $G_{0p}=G_{0n}=0.18$ MeV and $G_{2p}=G_{2n}=0.08$ MeV, respectively. 
The yrast and low-lying eigenstates are obtained as,
\begin{equation}
\left| \psi \right\rangle =\sum_{i}C_{i}\left| i\right\rangle \ ,
\label{eq:wf}
\end{equation}
where $\left\vert i\right\rangle $ is a cranked many-particle configuration and $C_{i}$ is the
corresponding probability amplitude. 

The angular momentum alignment $\left\langle J_{x} \right\rangle$ of
the state $\left\vert \psi \right\rangle$ is, 
\begin{eqnarray}
 \left\langle \psi \right| J_{x}
 \left| \psi \right\rangle
 = \sum_{i}\left|C_{i}\right| ^{2}
   \left\langle i\right| J_{x}\left| i\right\rangle
 + 2\sum_{i<j}C_{i}^{\ast }C_{j}
   \left\langle i\right| J_{x}\left| j\right\rangle.
 \label{eq:Jx1}
\end{eqnarray}
Since $J_{x}$ is an one-body operator, the matrix element $\left\langle i\right| J_{x}\left| j\right\rangle$ is nonzero only when $\left| i\right\rangle$ and $\left| j\right\rangle$ differ by one particle occupation, which are denoted by orbitals $\mu$ and $\nu$. Then $\left| i\right\rangle=(-)^{M_{i\mu}}\left| \mu\cdots\right\rangle$ and $\left| j\right\rangle=(-)^{M_{j\nu}}\left| \nu\cdots\right\rangle$ with the ellipsis stands for the same particle occupation and $(-)^{M_{i\mu}}=\pm1$ and $(-)^{M_{j\nu}}=\pm1$ according to whether the permutation is even or odd. The angular momentum alignment can be expressed as the diagonal and the off-diagonal parts,
\begin{eqnarray}
   \nonumber 
 \left\langle  J_{x} \right\rangle
& = &\langle J_x(\mu)\rangle+\langle J_x(\mu\nu)\rangle\\
   & =& \sum_{\mu} j_{x}(\mu)
 + \sum_{\mu<\nu} j_{x}(\mu\nu).
 \label{eq:Jx1}
\end{eqnarray}
The kinematic moment of inertia is given by $J^{(1)}=\left\langle \psi \right\vert J_{x}\left\vert\psi \right\rangle /\omega$. For the details of the PNC-CSM method, see Refs.~\cite{ZengJ1994_PRC50,ZengJ1983_NPA405,ZengJ1994_PRC50a}. 

\section{\label{sec:level3}CALCULATION AND DISCUSSION}

\begin{figure}[!t]
  \centering
  \includegraphics[width=4in]{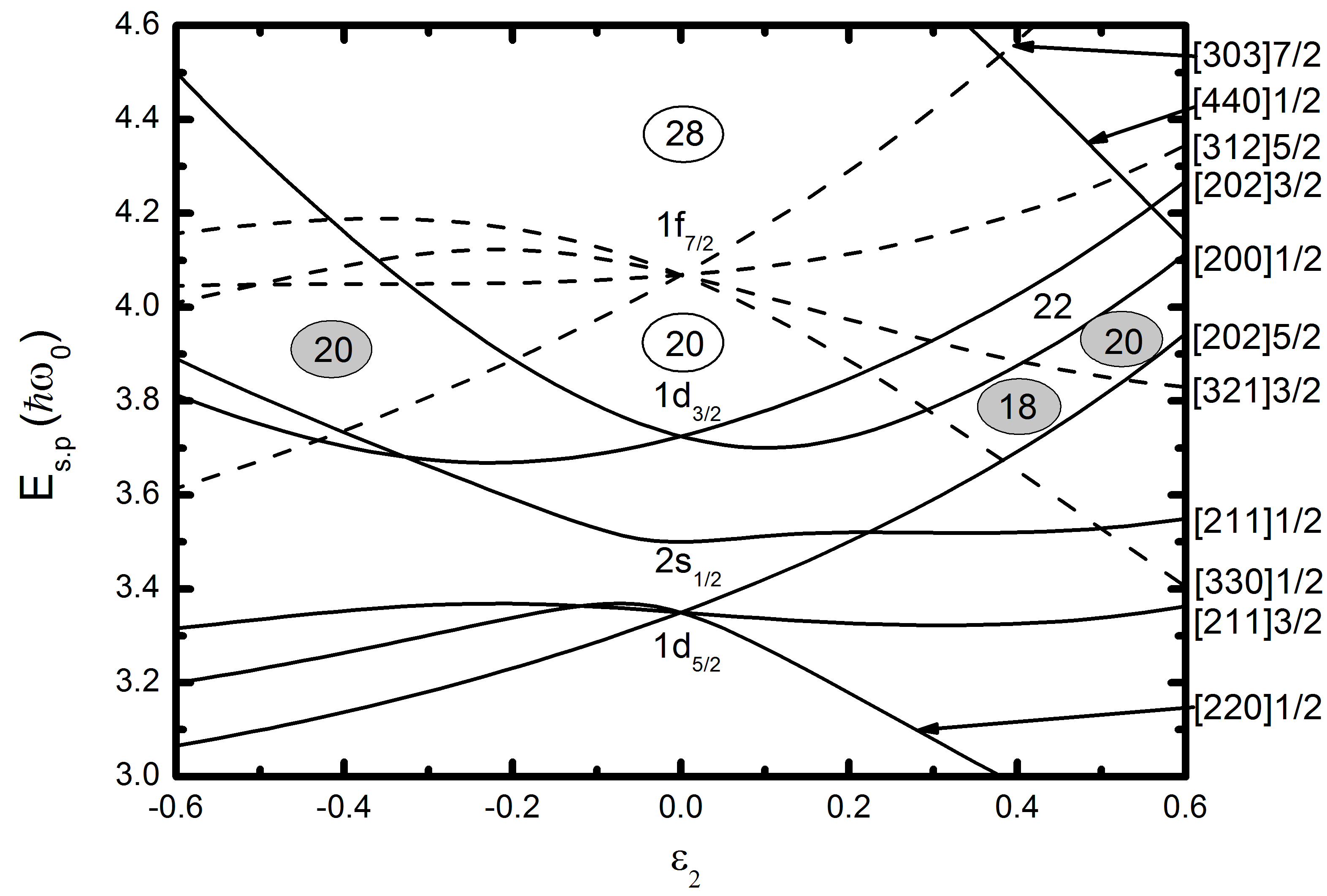}
  \caption{The Nilsson diagram for protons or neutrons around N=Z=20 with quadrupole deformation $\varepsilon_{2}$ ($\varepsilon_4=0$ and $\varepsilon_6=0$.). The Nilsson parameters $(\kappa,\mu)$ are taken from Ref.~\cite{BengtssonT1985_NPA436}. The $\kappa_{2},\kappa_{3}$ are modified slightly as $\kappa_{2}=0.08,\kappa_{3}=0.12$.}
  \label{Fig:fig1}
\end{figure}

The Nilsson parameters $(\kappa,\mu)$ are taken from Ref.~\cite{BengtssonT1985_NPA436}. Since $^{36}$Ar is a $N=Z$ symmetric nuclear system and the density of the single-particle level is low, the same set of $(\kappa,\mu)$ is used for neutrons and protons in the present calculations. The values of $\kappa_{2},\kappa_{3}$ are modified slightly to reproduce the correct single-particle level sequence. The corresponding Nilsson diagram for protons or neutrons is shown in Figure~\ref{Fig:fig1}. As it shows that the deformed $N=Z=18$ and $20$ energy gaps appear at the well-deformed prolate region around $\varepsilon_{2}=0.4$ and $0.5$, respectively. At the oblate deformation side, the $N=Z=20$ energy gap around $\varepsilon_{2}=-0.4$ is as large as the spherical shell gap at $N=Z=20$.    

The quadrupole deformation $\beta_2=0.45$ was suggested by cranked Nilsson-Strutinsky calculations for the SD bands of $^{36}$Ar in the original experimental paper~\cite{SvenssonC2000_PRL85}. Later, a large low spin quadrupole deformation $\beta_2=0.46\pm0.03$ is deduced from the $B(E2)$ value for $4^+\rightarrow2^+$ SD transition in Ref.~\cite{SvenssonC2001_PRC63}. As for $^{40}$Ar, the observed superdeformed structure was calculated by cranked Hartree-Fock-Bogoliubov with the $P + QQ$ force~\cite{IdeguchiE2010_PLB686}. The calculation shows that $\beta_{2}=0.57$ at $I = 0\hbar$ and the deformation gradually decreases to $0.45$ at $I = 12\hbar$. Triaxiality is found to be almost zero $(\gamma\approx 0^{\circ})$ throughout this angular momentum range. Calculations by the parity and angular momentum projection and the generator coordinate method suggest a quadrupole deformation $\beta_2=0.478$ and triaxial deformation with $\gamma\approx10^{\circ}$~\cite{TaniguchiY2010_PRC82}. In the PNC-CSM frame, nucleus is restricted to an axial symmetric shape in the whole spin range with the fixed deformation parameters. $\varepsilon_2=0.48$ and $\varepsilon_2=0.5$ are adopted for $^{36}$Ar and $^{40}$Ar, respectively. Higher order axial symmetric deformations of $\varepsilon_4=0.06$ and $\varepsilon_6=-0.06$ are included to reproduce the backbending of the SD band in $^{36}$Ar. 

\begin{figure}[!t]
  \centering
  \includegraphics[width=2.8in]{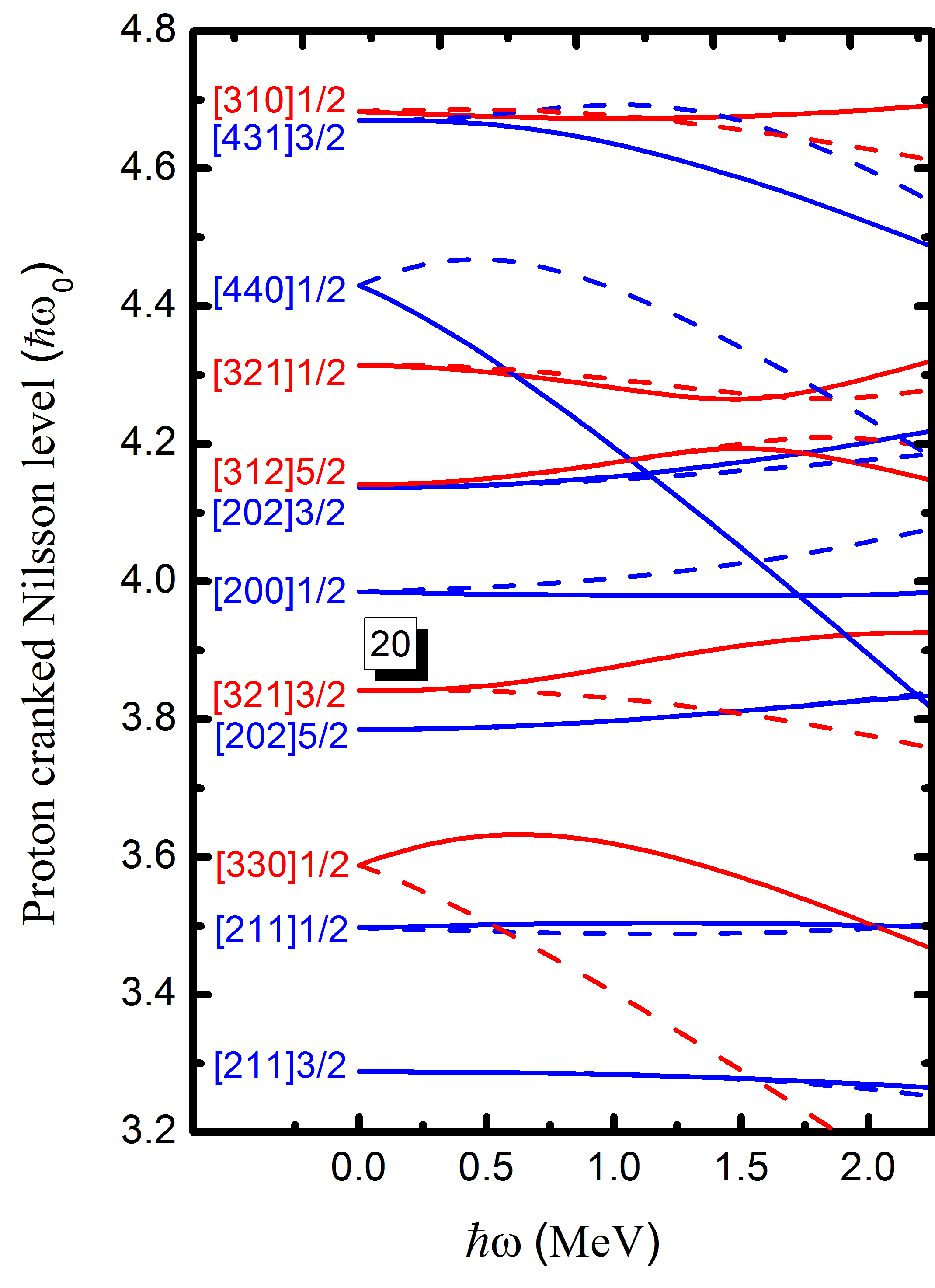}
  \caption{(colour online) Cranked proton Nilsson levels near the Fermi surface of $^{36}$Ar with quadrupole deformation parameter $\varepsilon_2=0.48$. The signature $\alpha=+1/2$ $(\alpha=-1/2)$ levels are denoted by solid (dash) lines. The positive (negative) parity levels are denoted by blue (red) lines. Cranked neutron Nilsson levels are the same.}
  \label{Fig:fig2}
\end{figure}

With the above selected parameters, the cranked Nilsson levels with quadrupole deformation parameter $\varepsilon_2=0.48$ for $^{36}$Ar are calculated in figure \ref{Fig:fig2}. It is same for neutrons and protons. The proton/neutron Fermi surface of $^{36}$Ar locates between the $1f_{7/2}[321]3/2$ and $1d_{5/2}[202]5/2$ orbitals. Since these two Nilsson orbitals stay close to each other and cross around $\hbar\omega=1.5$ MeV $(\alpha=-1/2)$, a band crossing is likely to arise around $\hbar\omega=1.5$ MeV. In contrast, the neutron Fermi surface of $^{40}$Ar is lifted up to the deformed shell gap at $N=22$ where the band crossing occurs between the $1d_{3/2}[200]1/2$ and $1g_{9/2}[440]1/2$ orbitals around $\hbar\omega=1.5$ MeV. Since $1g_{9/2}[440]1/2$ orbital is a high$-j$ low$-\omega$ intruder orbital, which is characterized by its large contributions to alignment and large Coriolis responses, a sharp backbending would arise around $\hbar\omega=1.5$ MeV. However, the experimentally observed SD band in $^{40}$Ar is up to spin $I^{\pi}=12^{+}$ which is equivalent to $\hbar\omega=1.35$ MeV. Then the predicted band crossing would occur beyond the experimental observed frequency range. Therefore, we will not discuss about it.  

\begin{figure}[!t]
  \centering
  \includegraphics[width=4in]{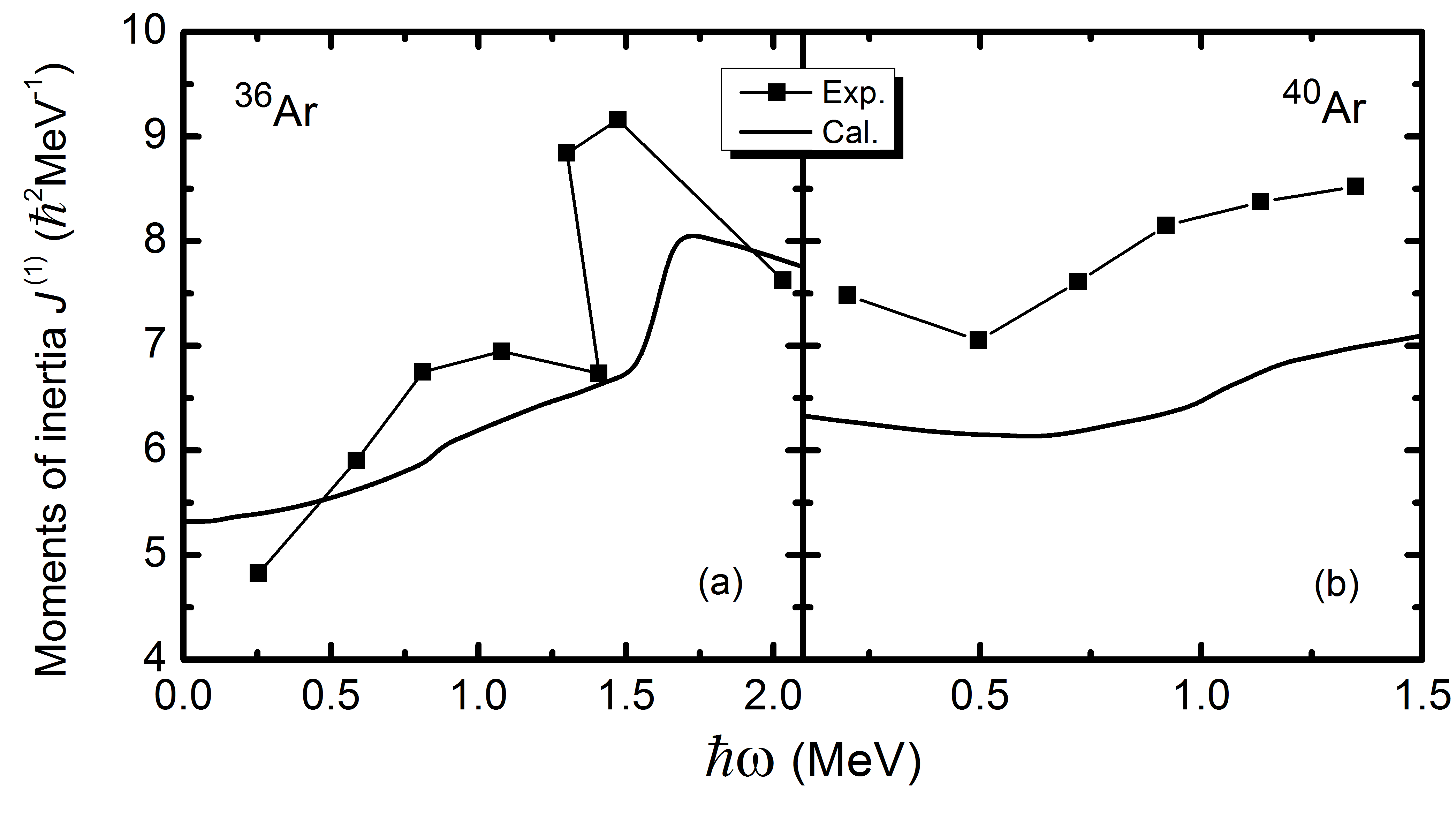}
  \caption{The comparison of the experimental kinematic moment of inertia $J^{(1)}$ of SD bands in $^{36}$Ar (a) and $^{40}$Ar (b) with the PNC-CSM calculations. Experimental data are denoted by solid squares and theoretical results are denoted by solid lines.}
  \label{Fig:fig3}
\end{figure}

The comparison of the theoretical $J^{(1)}$ with the extracted experimental values for SD bands in $^{36,40}$Ar is plotted in Figure~\ref{Fig:fig3}. The near-perfect rotational behavior of the $^{36}$Ar was observed in experiment up to spin $I=10\hbar$ (around rotational frequency $\hbar\omega=1.5$ MeV) where a backbending arises~\cite{SvenssonC2000_PRL85}. The agreement of the backbending frequency around $\hbar\omega=1.5$ MeV is remarkably good.  The calculated backbending of $J^{(1)}$ at $\hbar\omega>1.5$ MeV is less pronounced than the experimental data. The Cranked Nilsson-Strutinsky calculation shows the system maintains an axially symmetric shape before the backbending while it changes the shape to have the triaxial deformation after the backbending~\cite{SvenssonC2001_PRC63}. The cranked Skyrme-Hartree-Fock calculation reveals that the shape of the superdeformed $^{36}$Ar system becomes triaxial and evolves toward the oblate shape at high spin limit. The PNC-CSM calculation is carried out with the fixed symmetric deformations throughout the whole observed frequency range. To develop present model to take into account the shape evolution may improve quantitive agreement between theoretical results and experimental data.  

\begin{figure}[!t]
  \centering
  \includegraphics[width=4in]{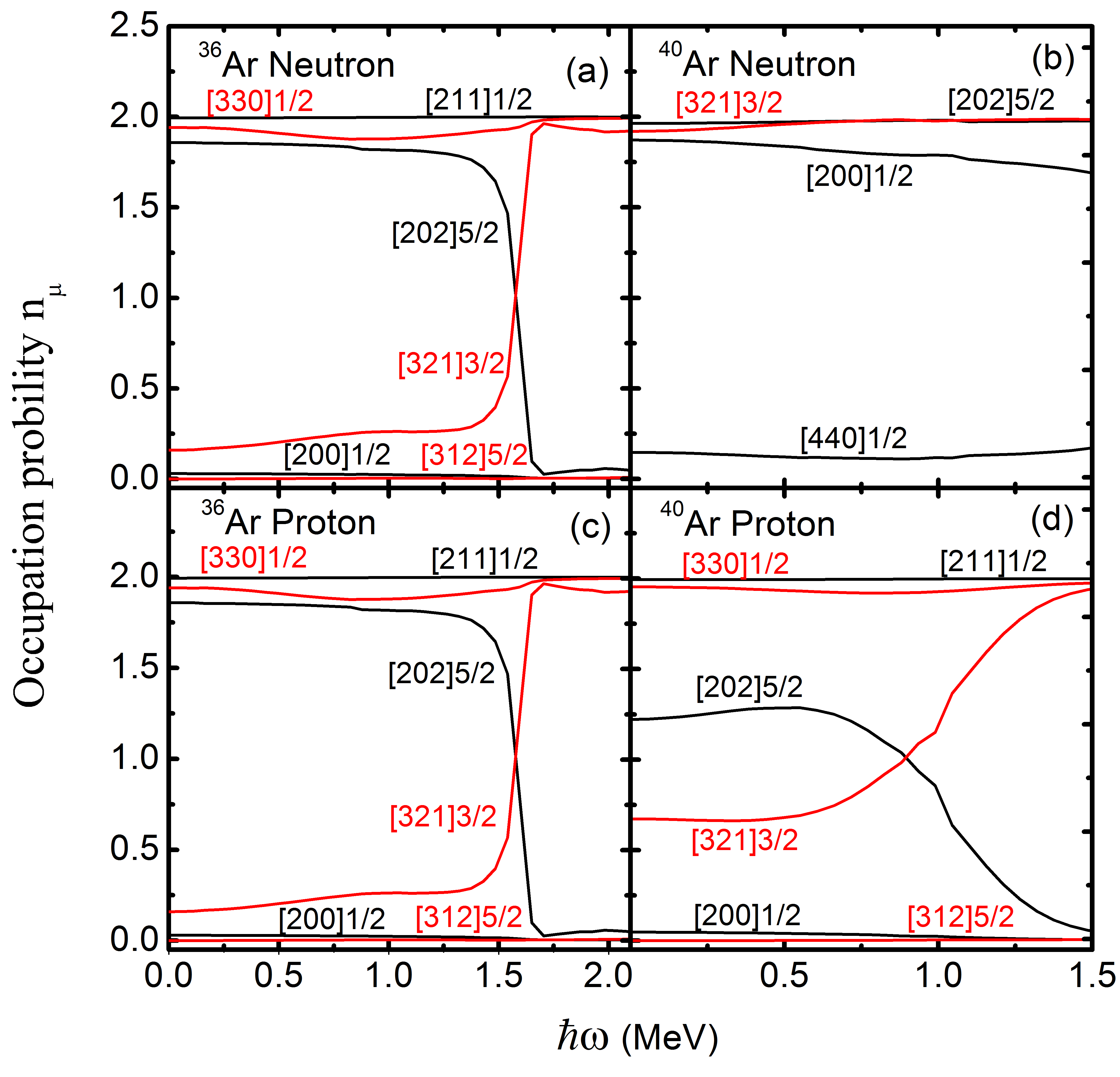}
  \caption{  (colour online) Occupation probability $n_{\mu}$ of each orbit $\mu$ (including both $\alpha=\pm1/2$) near the Fermi surface for SD bands in $^{36}$Ar (left) and $^{40}$Ar (right). $n_{\mu}$ of the positive (negative) parity levels are denoted by black (red) lines. }
  \label{Fig:fig4}
\end{figure}

To reveal the microscopic mechanism of the backbending of the rotational band is often interested in theoretical studies since it can provide valuable informations to understand deeply the microscopic structures of the rotating nuclear system. Based on the analysis of projected shell model calculation, the backbending of the SD band in $^{36}$Ar is explained as the results of the 0-, 2-, and 4-quasiparticle (qp) bands cross each other at the same angular momentum $I^{+}=10^{+}$~\cite{LongG2001_PRC63}.

In the PNC-CSM calculations, the band crossing in $^{36}$Ar is clearly exhibited by the occupation probabilities $n_{\mu}$ of each cranked Nilsson orbit $\mu$ in figure~\ref{Fig:fig4}(a) and~\ref{Fig:fig4}(c). We can see that $n_{\mu}$ of neurons and protons are the same. Before the backbending (at $\hbar\omega\le1.5$ MeV), $1f_{7/2}[321]3/2$, orbital just above the Fermi surface, is almost empty $(n_{\mu}\approx0)$ and $1d_{5/2}[202]5/2$, orbital just below the Fermi surface, is almost fully occupied $(n_{\mu}\approx2)$ whereas it exchanges after the backbending. Therefore, the backbending results from the simultaneous band crossing of neutrons and protons between the ground state (0-qp) band and the $1f_{7/2}[321]3/2$ (with signature $\alpha=\pm1/2$) configuration state (4-qp) band. This is consistent with the conclusion of projected shell model calculations. Furthermore, the PNC-CSM calculations present clearly why in contrast to the common band crossing picture, the 2-qp configurations do not have a chance to play a major role in the structure of the SD yrast band in $^{36}$Ar. Since $^{36}$Ar is a $N=Z$ symmetric nucleus, the neutron and proton signature pairs are excited from $1d_{5/2}[202]5/2$ to $1f_{7/2}[321]3/2$ configuration state simultaneously to form a 4-qp state immediately after the backbending. 

\begin{figure}[!t]
  \centering
  \includegraphics[width=4in]{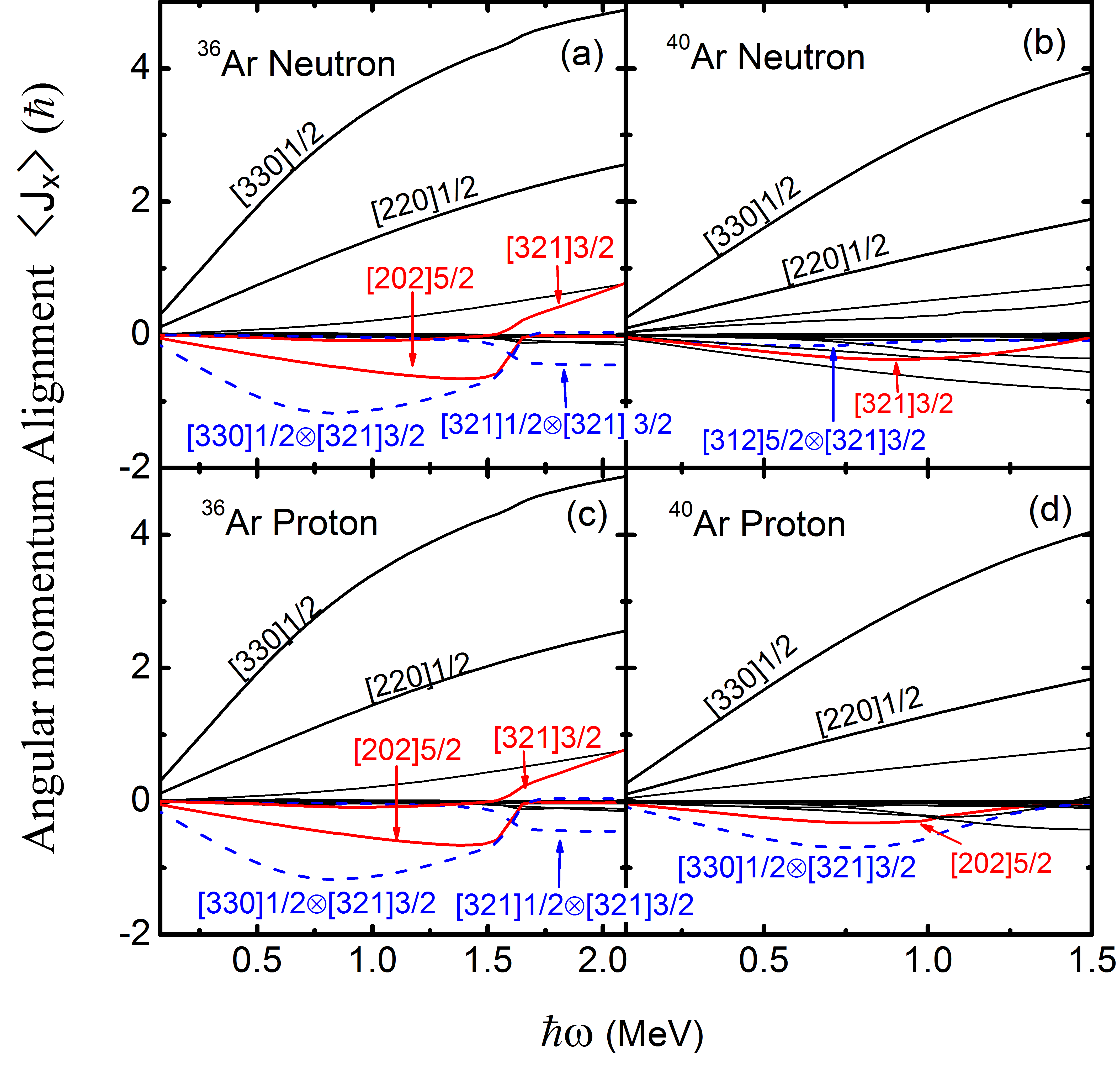}
  \caption{ (colour online) The direct contributions to the angular momentum alignment $\langle J_x\rangle$  from the particle occupying the cranked orbit $\mu$ (denoted by solid lines) and the interference $\langle J_x(\mu\nu)\rangle$ between orbit $\mu$ and $\nu$ (denoted by the dashed lines) for the SD band of $^{36}$Ar (left) and $^{40}$Ar (right).}
  \label{Fig:fig5}
\end{figure}

To quantify the effect, more detailed informations of the angular momentum alignment $\langle J_x(\mu)\rangle$ from each single particle orbit $\mu$ and the interference $\langle J_x(\mu\nu)\rangle$ between orbit $\mu$ and $\nu$ are presented in figure~\ref{Fig:fig5}. As it shows in figure~\ref{Fig:fig5} (a) and~\ref{Fig:fig5} (c), the sharp rise of the $J^{(1)}$ in $^{36}$Ar mainly results from the contribution of the sudden increased simultaneous alignments of neutron and proton $1d_{5/2}[202]5/2$ pairs and $1f_{7/2}[321]3/2$ pairs at $\hbar\omega=1.5$ MeV. Besides, their involved interference terms are important too. While $\langle J_{x}([321]1/2\otimes[321]3/2)\rangle$ decreases suddenly at $\hbar\omega=1.5$ MeV, $\langle J_{x}([330]1/2\otimes[321]3/2)\rangle$ shows an excessive sharp increase. The smooth ascend of the $J^{(1)}$ at the low frequency is attributed to the gradual alignments of neutron and proton $1f_{7/2}[330]1/2]$ pairs and $1d_{5/2}[220]1/2$ pairs.        

The rotational behavior of the SD band in $^{40}$Ar is quite different. Only a slight upbending of $J^{(1)}$ appears at the low spin (around $\hbar\omega=0.5$ MeV). Then $J^{(1)}$ increases smoothly with rotational frequency. The PNC-CSM calculations reproduce the experimental variation tendency very well. However, the theoretical results underestimate the data by about $1.2$ $\hbar^{2}$MeV$^{-1}$ throughout the whole observed frequency range. The $n-p$ pairing would be important in such (near-)symmetric nuclear system, that is not included in the present PNC-CSM method. This could be one of the reasons for the systematic shift down of the theoretical results. Nevertheless, further investigations aimed at the effect of the $n-p$ pairing should be fulfilled by the PNC-CSM method in this mass region. 

Effected by the four additional neutrons in $^{40}$Ar, the rotational behavior differs a lot with that of $^{36}$Ar. From the occupation probabilities in figure~\ref{Fig:fig4} (b), we can see that the $1d_{3/2}[200]1/2$ and $1f_{7/2}[321]3/2$ orbitals are fully occupied $(n_{\mu}\approx2)$. Since the deformed $N=22$ shell gap is comparatively big, there is no neutron band crossing occurring during the experimentally observed frequency range. The proton occupation probabilities [see figure~\ref{Fig:fig4} (d)] are effected accordingly due to the change of the mean field. $1d_{5/2}[202]5/2$ orbital is more than half occupied $(n_{\mu}\approx1.25)$, and $1f_{7/2}[321]3/2$ orbital is less than half occupied $(n_{\mu}\approx0.75)$. From figure~\ref{Fig:fig5} (b) and~\ref{Fig:fig5} (d), the slight upbending at the low frequency is attributed to the alignments of the neutron $1f_{7/2}[321]3/2$ pairs and proton $1d_{5/2}[202]5/2$ pairs, and their involved interference terms of neutron $\langle J_{x}([312]5/2\otimes[321]3/2)\rangle$ and proton $\langle J_{x}([330]1/2\otimes[321]3/2)\rangle$. 

\section{\label{sec:level4}CONCLUSIONS}
For the first time the cranked shell model with the paring correlations treated by the particle-number-conserving method has been used to describe the superdeformed rotational bands in the $A=40$ mass region. The calculations are carried out within $N=0\sim4$ major shells, with axially symmetric deformation parameters $\varepsilon_{2,4,6}$ being included and the pairing  correlations being treated properly. The experimental kinematic moments of inertia $J^{(1)}$ versus rotational frequency in $^{36}$Ar and $^{40}$Ar are reproduced well. This may convince us that the PNC-CSM method is an efficient method to describe the rotational properties of the superdeformed nuclei around the $A=40$ mass region. 

The microscopic mechanism of the variation of the superdeformed bands versus frequency is explicit in the PNC-CSM calculations. The backbending around $\hbar\omega=1.5$ MeV of the SD band in $^{36}$Ar is clearly presented by analysis the dominant components of the total wave function of the cranked shell model Hamiltonian. It is attributed to the simultaneous alignments of neutron and proton $1d_{5/2}[202]5/2$ pairs and $1f_{7/2}[321]3/2$ pairs, which is caused by the band crossing between the $1d_{5/2}[202]5/2$ and $1f_{7/2}[321]3/2$ configuration states. As for $^{40}$Ar, four more additional neutrons rise the neutron Fermi surface to the $N=22$ deformed shell gap. There is no band crossing occurring during the experimentally observed frequency range. Therefore the variation of the $J^{(1)}$ versus frequency is much gentle. The slight upbending at the low frequency is mainly effected by the alignments of the neutron $1f_{7/2}[321]3/2$ pairs and proton $1d_{5/2}[202]5/2$ pairs. Moreover, the PNC-CSM results show that besides the diagonal parts, the off-diagonal parts is very important. We see that not only $1f_{7/2}$ orbital, but also $1d_{5/2}$ plays a very important role in the rotational behavior of the SD bands in $^{36,40}$Ar, which can not be neglected. 

\section{\label{sec:level5}ACKNOWLEDGMENTS}
This work was supported by the National Natural Science Foundation of China under Grant No. 11775112 and 11275098, and the Priority Academic Program Development of Jiangsu Higher Education Institutions.

\bibliographystyle{unsrt.bst}
\bibliography{../../../../References/ReferencesXT}

\end{document}